\documentclass[11pt,a4paper]{article}
\usepackage{jheppub}
\pdfoutput=1
\usepackage{amsfonts,amssymb,amsmath}
\usepackage{graphicx,epstopdf}% http://ctan.org/pkg/graphicx
\usepackage{array}% http://ctan.org/pkg/array
\usepackage{placeins}
\usepackage{slashed}
\usepackage{youngtab}
\usepackage{physics}

\usepackage{IEEEtrantools}
\usepackage[color,final]{showkeys} %%% should be loaded after hyperref        %%% add final to avoid printing the labels
\newcommand{\beq}{\begin{equation}}

\newcommand{\eeq}{\end{equation}}

\newcommand{\nn}{\nonumber}

\newcommand{\mD}{\mathcal{D}}

\newcommand{\mN}{\mathcal{N}}

\newcommand{\mS}{\mathcal{S}}

\newcommand{\p}{\partial}

\newcommand{\f}{\frac}
\newcommand{\half}{\frac{1}{2}}

\newcommand{\al}{\alpha}
\newcommand{\be}{\beta}
\newcommand{\ga}{\gamma}         
\newcommand{\de}{\delta}        \newcommand{\De}{\Delta}
\newcommand{\ep}{\epsilon}

\newcommand{\te}{\theta}

\newcommand{\ka}{\kappa}
\newcommand{\la}{\lambda}       

\newcommand{\rh}{\rho}

\newcommand{\ph}{\phi}          
\newcommand{\ps}{\psi}

\def\pr{\partial}
\newcommand{\bph}{\bar{\ph}}

\newcommand{\bps}{\bar{\psi}}

\begin{document}

\title{Dual Superconformal Symmetry of ${\cal N}=2$ Chern-Simons theory with Fundamental Matter at Large $N$}
\author[a]{Karthik Inbasekar,}
\author[b]{Sachin Jain,}
\author[b]{Sucheta Majumdar,}
\author[c]{Pranjal Nayak,}
\author[b]{Turmoli Neogi,}
\author[d]{Ritam Sinha,}
\author[e]{Tarun Sharma,}
\author[e]{V Umesh}

\affiliation[a]{Faculty of Exact Sciences, School of Physics and Astronomy, Tel Aviv University, Ramat Aviv 69978, Israel}
\affiliation[b]{Indian Institute of Science Education and Research, Homi Bhabha Rd, Pashan, Pune 411 008, India}
\affiliation[c]{Department of Physics \& Astronomy, 265 Chemistry-Physics Building, University of Kentucky, Lexington, 40506, USA}
\affiliation[d]{Department of Theoretical Physics, Tata Institute of Fundamental Research, Navy Nagar, Mumbai 400005, India}
\affiliation[e]{National Institute for Theoretical Physics, School of Physics and Mandelstam Institute for Theoretical Physics, University of the Witwatersrand, Johannesburg Wits 2050, South Africa}

\emailAdd{karthikin@tauex.tau.ac.il}
\emailAdd{sachin.jain@iiserpune.ac.in}
\emailAdd{sucheta.majumdar@students.iiserpune.ac.in}
\emailAdd{nayak.pranjal@gmail.com}
\emailAdd{turmoli.neogi@students.iiserpune.ac.in}
\emailAdd{tarunks21@gmail.com}
\emailAdd{ritamsinha.physics@gmail.com}
\emailAdd{vumesh.physics@gmail.com}

\abstract{
Dual conformal symmetry and Yangian symmetry are symmetries of amplitudes that have aided the study of scattering amplitudes in highly supersymmetric theories like ${\cal N}=4$ SYM and ABJM. However, in general such symmetries are absent from the theories with lesser or no supersymmetry. In this paper, we show that the tree level $2\to 2$ scattering amplitude in the 3d ${\cal N}=2$ Chern-Simons theory coupled to a fundamental chiral multiplet is dual superconformal invariant. In the 't Hooft large $N$ limit, the $2\to 2$ scattering amplitude in this theory has been shown to be tree-level exact in non-anyonic channels, while having only an overall multiplicative coupling dependent renormalisation in the anyonic channel. Therefore, the dual superconformal symmetry that we demonstrate in this paper is all loop exact. This is unlike the previously studied highly supersymmetric theories where dual superconformal symmetry is anomalous at loop levels.

Furthermore, we reverse the argument to study the extent to which dual superconformal invariance fixes the scattering amplitude in an ${\cal N}=2$ supersymmetric theory. We demonstrate that requiring the dual superconformal invariance completely fixes the momentum dependence of the $2\to2$ amplitude, while the coupling constant dependence remain unfixed. Further, we use a combination of parity invariance, unitarity and self-duality of the amplitude to constrain the coupling dependence of scattering amplitude.
}

\begin{flushright}
TAUP-3027/17,TIFR/TH/17-42 
\end{flushright}

\maketitle  

\section{Introduction}\label{intro}
Scattering amplitudes are one of most basic and interesting observables in quantum field theories. The usual method of computing these amplitudes via the Feynman diagrams, though very useful, is usually restricted only to low orders in perturbation theory and small number of scattering particles as the computational difficulty rapidly increases with the number of loops and particles involved. 

The study of scattering amplitudes in highly (super) symmetric QFTs have revealed that Feynman diagrams are perhaps not the best way to compute scattering amplitudes.  Over the past decade or so, a lot of progress has been made in alternative ways to compute amplitudes that utilize hidden symmetry structures not visible directly from the Lagrangian \cite{Drummond:2008vq, Alday:2010zy, Drummond:2009fd}, both for tree amplitudes as well as for efficient loop computations (see \cite{Elvang:2015rqa} and reference therein for more details). 

In such endeavors it is useful to have models where one can compute observables exactly. Such examples serve as a useful ground for development and testing of new techniques and ideas. Though such models may or may not have a direct connection to intended real world physics, nevertheless, they can provide crucial insights into the structural features for similar observables in more realistic models. 4d ${\cal N}=4$ SYM and 3d ${\mathcal N}=6$ ABJM theory are well known examples of such models. These theories have some remarkable properties. They have weakly coupled\footnote{In 't Hooft limit of the field theory.} holographic duals that are extremely useful for computing corresponding observables in the strong coupling limit of the field theory. Scattering amplitudes in these models show some unusual symmetries, such as Dual Superconformal symmetry and infinite dimensional Yangian symmetry, that have led to remarkable developments such as an Orthogonal Grassmanian representation for the scattering amplitude \cite{ArkaniHamed:2012nw}. 

Since both these theories are highly supersymmetric, it is natural to ask if these features are necessarily tied to the high amount of supersymmetry or can one have theories with lesser, or perhaps, no supersymmetry and still retain some, or all, of these nice features of scattering amplitudes?

In this paper, we study a theory with much less supersymmetry: the ${\cal N}=2 ~~U(N)$ Chern-Simons (CS) theory coupled to matter in fundamental representation in the large $N$ limit, and demonstrate that it possesses many of the above mentioned symmetries of the amplitudes. Chern-Simons theories coupled to fundamental matter have been a subject of intense research \cite{Aharony:2011jz,Giombi:2011kc,Maldacena:2011jn,Maldacena:2012sf,Chang:2012kt,Aharony:2012nh,Jain:2012qi,Yokoyama:2012fa,GurAri:2012is,Aharony:2012ns,Jain:2013py,Takimi:2013zca,Jain:2013gza,Yokoyama:2013pxa,Jain:2014nza,Gurucharan:2014cva,Dandekar:2014era,Frishman:2014cma,Aharony:2015pla,Bedhotiya:2015uga,xyz,Minwalla:2015sca,Geracie:2015drf,Wadia:2016zpd,Yokoyama:2016sbx,Giombi:2016ejx,Radicevic:2016wqn,Gur-Ari:2016xff,Giombi:2016zwa,Giombi:2017txg,Nosaka:2017ohr} over the past few years and via a series of impressive all loop exact computations of a variety of observables lots of non trivial evidence has been gathered for a Strong-Weak Bose-Fermi duality. These models show other interesting features such as the existence of weakly broken higher spin symmetry and holographic duality with Vasiliev higher spin theories in $AdS_4$. Apart from these formal considerations, the finite $N$ version of these theories also have relevance for condensed matter, namely quantum hall systems. The large $N$ Bose-Fermi dualities in these theories have later also been conjecturally generalised \cite{Radicevic:2015yla,Aharony:2015mjs,Seiberg:2016gmd,Karch:2016sxi,Karch:2016aux,Hsin:2016blu,Aharony:2016jvv,Benini:2017dus,Gaiotto:2017tne,Jensen:2017dso,Jensen:2017xbs} to finite $N$. These theories are easily generalised to include supersymmetry where the Strong-Weak Bose-Fermi duality is related to the known supersymmetric  dualities proposed in \cite{Benini:2011mf}, generalising the Giveon-Kutasov duality \cite{Giveon:2008zn}. 

In \cite{Inbasekar:2015tsa}, an all loop exact computation of the $2 \to 2$ scattering amplitudes in ${\cal N}=1$ and ${\cal N}=2$ supersymmetric version of these theories with a single chiral multiplet was performed. Quite surprisingly, it turned out that, in the large $N$ limit, the $2 \to 2$ scattering amplitude in ${\mathcal N}=2$ theory does not receive loop corrections, except in the anyonic channel.\footnote{Following \cite{Jain:2014nza}, we refer the gauge singlet exchange channel in the particle-antiparticle scattering as the anyonic channel. As explained in \cite{Jain:2014nza}, the crucial feature in this scattering channel is that the Aharonov-Bohm phase, which imparts anyonic nature (fractional statistics) to the scattering particles, survives in the large $N$ 't Hooft limit in this channel.}\textsuperscript{,}\footnote{Since the amplitude does not receive correction in the non-anyonic channel due to the loop diagrams, the momentum dependence of the full amplitude is same as that of the tree-level amplitude.} In the anyonic channel also the momentum dependence of the amplitude stays the same as that of the tree-level amplitude and the quantum correction only gives rise to a multiplicative coupling dependent term. 
We wish to understand what symmetries, if any, directly give rise to such remarkable simplicity of the amplitude in this theory and propose dual superconformal symmetry as a possible candidate. Further, it was recently shown in \cite{Inbasekar:2017ieo}, that all the tree level scattering amplitudes can be computed using BCFW recursion relation for ${\cal N}=2$ theory. This makes higher-point scattering amplitudes in this theory amenable to the study of these symmetries, something that we leave for future work.

In this paper, by explicitly performing the check on the exact result of \cite{Inbasekar:2015tsa}, we show that the $2 \to 2$ scattering amplitude in ${\cal N}=2$ theory is {\it dual superconformal invariant} to all loop order at leading order in the ${1}/{N}$ expansion.  Firstly let us emphasize that it is not obvious that the amplitude in the $\mathcal{N}=2$ theory is dual superconformal invariant. The product of the momentum and supermomentum conservation delta functions is not invariant under dual inversions unlike that of the amplitudes in $\mN=4$ SYM and ABJM. To the best of our knowledge this is the first time a claim has been made that the amplitude in the $\mathcal{N}=2$ Chern-Simons matter theory is dual superconformal invariant. All these developments strongly suggest that along with ${\mathcal N}=4$ SYM in 4d, ${\mathcal N}=2$ CS theory with fundamental representation, might turn out to be an excellent model to many new developments for modern techniques in scattering amplitude computation.
%%%%%%%%%%%%%%%%%%%%%%%%%%%%%%%%%

%%%%% ------ This line (with modification) should move to the discussion/conclusion section ----  %%%%%%%
%We believe that all these properties makes ${\mathcal N}=2$ CS theory a model to discuss and develop the new tools in S-matrix computation, on par with  ${\mathcal N}=4$ Super Yang-Mills theory in four dimensions and ${\mathcal N}=6$ ABJM Chern-Simons theory in three dimensions.

This paper is organized as follows. In \autoref{scat}, we begin with a brief summary of scattering amplitudes in ${\cal N}=2$  CS matter theory. In \autoref{sec:dsc}, we introduce the dual coordinates and show the Dual Superconformal invariance of the scattering amplitude at all loop. In \autoref{sec:ulta}, we reverse the argument and ask, is it possible to fix the amplitude completely based on general principles? We show, dual superconformal invariance fixes the momentum dependent piece leaving the coupling constant dependence unfixed. In \autoref{subsec:parity-unitarity-duality}, we impose parity invariance, unitarity and self-duality of the amplitude to examine the constraints on the coupling constant dependence of the scattering amplitude. In \autoref{sec:cft}, using the fact that, in dual space the scattering amplitude can be thought of as correlation function of some operators, we show that full amplitude can be accounted for just by an identity operator exchange in one of the OPE channel of the correlator, which implies that the dual CFT is free. We end in \autoref{sec:diss} with discussion of the results and a list of possible future directions.

%%%%%%%%%%%%%%%%%%%%%%%%%%%%%%%%%%%%%%%%%%%%%%%%
\section{Amplitude in \texorpdfstring{$\mN=2$}{N=2} Chern-Simons matter theory}\label{scat}
In this section, we start with a brief review of the results for scattering amplitude computations performed in \cite{Inbasekar:2015tsa}. We will mainly focus on ${\mathcal N}=2$ $SU(N)$ Chern-Simons theory at level $\kappa$ coupled to a single chiral multiplet in fundamental representation. The action for the theory is 
\begin{align}\label{susycs}
\mS_{\mN=2}^L = \int d^3x 
\biggl[&-\f{\ka}{4\pi}\ep^{\mu\nu\rh}\text{Tr}\left( A_\mu\p_\nu A_\rh-\f{2i}{3}A_\mu A_\nu A_\rh\right)+\bps i \slashed{\mD}\ps-\mD^\mu\bph\mD_\mu\ph\nn\\
&+\f{4\pi^2}{\ka^2}(\bph\ph)^3+\f{4\pi}{\ka}(\bph\ph)(\bps\ps)+\f{2\pi}{\ka}(\bps\ph)(\bph\ps)\biggr]\ .
\end{align}
This theory has been conjectured \cite{Benini:2011mf} to have a strong-weak type self duality. The duality transformation on parameters, in a 't Hooft like large $N$ limit of $\kappa,N \rightarrow \infty$ with $\la = \frac{N}{\kappa}$ fixed, are 
\begin{equation}\label{dualityst}
\kappa=-\kappa, \quad \lambda=\la-\text{sgn}(\lambda)\ .
\end{equation} 

To study the scattering amplitudes, it is convenient to introduce the spinor helicity basis \cite{Elvang:2015rqa}, defined by
\beq
\begin{split}
p_i^{\alpha\beta} & =p_i^{\mu}\sigma_{\mu}^{\alpha\beta}=\lambda_i^{\alpha}\lambda_i^{\beta}, \\
(p_i+p_j)^2 & = 2 p_i.p_j=-\langle \lambda_i^{\alpha} \lambda_{j,\alpha}\rangle^2 . 
\end{split}
\eeq
Henceforth, we use the notation $\langle i j\rangle \equiv \langle\lambda_i^{\alpha} \lambda_{j\alpha}\rangle.$
The two on-shell supercharges for $n$ point scattering amplitudes  are given by
\beq
\begin{split}\label{SChrg}
{\mathcal Q} & =\sum_{i=1}^{n}q_i =\sum_{i=1}^{n}\lambda_i \eta_i, \\ 
{\bar {\mathcal Q}} & =\sum_{i=1}^{n}{\bar q}_i =\sum_{i=1}^{n}\lambda_i \partial_{\eta_i} .
\end{split}
\eeq
where $\eta$ is the on-shell spinor variable (see section II of \cite{Inbasekar:2017ieo} for more details).

\paragraph{Tree level $2\rightarrow 2$ scattering:} The scattering amplitude can be analyzed into two categories, namely anyonic and non-anyonic. To explain this, let us consider scattering of two particles, one in fundamental and another in anti fundamental representation of $U(N)$. This scattering can happen through two channels, namely adjoint and singlet (see \cite{Jain:2014nza} for details). Similarly, we can consider scattering of two particles, both in fundamental representation. The intermediate channel would be symmetric or anti-symmetric channel. We call, scattering through singlet channel to be anyonic where as rest of the channels are called non-anyonic channel. As was shown in \cite{Inbasekar:2015tsa}, at tree level one can go from anyonic to non-anyonic channel using usual crossing symmetry. However, at the loop one needs to modify the crossing relation (see \cite{Jain:2014nza,Inbasekar:2015tsa} for details). The tree level $2 \to 2$ super-amplitude in this theory is given by \cite{Inbasekar:2017ieo}
\begin{equation}\label{eq:super4pt}
T_{\rm {tree}}=\frac{4\pi}{\kappa}\sqrt{\frac{\langle 1 2\rangle^2}{\langle 2 3\rangle^2}}\delta(\sum_{i=1}^{4}p_i) ~\delta^2\left({\mathcal Q}\right) = \frac{4\pi}{\kappa} \mathcal A_4.
\end{equation}

\paragraph{Exact loop level $2\rightarrow 2$ scattering:} Surprisingly, it was shown in \cite{Inbasekar:2015tsa} that in the large $N$ limit the $2 \to 2$ scattering amplitudes in this theory do not get renormalized by loop corrections except in the anyonic channel where the quantum corrections lead only to a very simple overall coupling dependence. The amplitudes are given by
\begin{align}\label{loopans1}
T_{{\rm all-loop}}^{{\rm non-anyonic}} &=T_{\rm {tree}} \\
T_{{\rm all-loop}}^{{\rm anyonic}} &=N \frac{\sin(\pi\lambda)}{\pi\lambda}T_{\rm {tree}}. \label{loopans2}
\end{align}
The full $2 \to 2$ $S$-matrix takes the very simple form 
\begin{equation}\label{S-matrix-a}
\begin{split}
S^{{\rm non-anyonic}}&=I+ i ~T_{{\rm all-loop}}^{{\rm non-anyonic}}\\
S^{{\rm anyonic}}&=\cos\left(\pi\lambda\right)I + i ~T_{{\rm all-loop}}^{{\rm anyonic}}.
\end{split}
\end{equation}
Note that in the anyonic channel, consistency with unitarity demands that the forward scattering amplitude, $I$, also gets renormalized. This renormalisation of forward scattering is a reflection of the anyonic phase as explained in \cite{Jain:2014nza,Inbasekar:2015tsa}. 
 
This exceptional simplicity of the exact $2\to 2$ S-matrix in this theory begs for a symmetry based explanation. The main aim of the current paper is to make progress in this direction based on an exact loop level dual superconformal symmetry. 

\section{Dual Superconformal symmetry}\label{sec:dsc}
Dual superconformal invariance has played a crucial role in computation of scattering amplitudes both in ${\mathcal N}=4$ SYM as well as in ${\mathcal N}=6$ ABJM theory. This has led to new developments like Yangian invariant Grassmanian representation. Both of these examples are highly supersymmetric and it was commonly believed that dual superconformal invariance may be related to high amount of  supersymmetry. In this section we will show that, in spite of much less supersymmetry, namely ${\mathcal N}=2$, the theory has dual superconformal invariance.  This symmetry is nothing but superconformal invariance of the scattering amplitude when expressed in the dual variables, that we introduce shortly. As we will see, this symmetry is not manifest either in usual momentum variables or in position superspace variables. We begin our discussion with the definition of the dual space coordinates and dual superconformal transformation in the next two subsections and later show the invariance of the four point amplitude in \eqref{eq:super4pt}.

%%%%%%%%%%%%%%%%%%%%%%%%%%%%%%%%%%%%%%
\subsection{Dual space}
Following \cite{Drummond:2008vq, Gang:2010gy}, we define the dual space co-ordinates $(x,\te)$ as
\beq
\begin{split}\label{Ceqns}
& x_{i,i+1}^{\al\be} = x_i^{\al\be}-x_{i+1}^{\al\be} = p_i^{\al\be}=\la_i^{\al}\la_i^{\be} \\
& \te_{i,i+1}^{\al} = \te_i^{\al}-\te_{i+1}^{\al} = q_i^{\al}=\la_i^{\al}\eta_i  \\
\end{split}
\eeq
where $i=1,\dots,n$. We also impose the conditions that $x_{n+1}=x_1$ and $\te_{n+1} = \te_1$. These condition imply
\beq
\begin{split}\label{pqcons}
P^{\al\be} & = \sum_i p_i^{\al\be}= x_{n+1}^{\al\be} - x_1^{\al\be} = 0 , \\
{\mathcal Q}^{\al} & = \sum_i q_i^{\al} = \te_{n+1}^\al - \te_1^\al = 0 .
\end{split}
\eeq
In the dual space, we wish to show the existence of an additional symmetry that is manifest at the level of the amplitudes, namely the {\it Dual Superconformal
Symmetry}. To show this, let us start by defining its action on the dual coordinates $\{ x_i^{\al\be}, \te_i^\al \}$.

%%%%%%%%%%%%%%%%%%%%%%%%%%%%%%%%%%%%%%%
\subsection{Action of dual superconformal generators}
As we will show, the dual superconformal symmetry in the amplitudes \eqref{eq:super4pt} will form the supergroup $osp(2|4)$\footnote{This group is same as the position-space superconformal symmetry group of this theory.}. We will denote the dual superconformal generators with upper case letter 
\beq 
\{ P_{\al\be}, ~M_{\al\be}, ~D, ~K_{\al\be}, ~R, ~Q_\al, ~\bar Q_\al, ~S_\al, ~\bar S_\al \}
\eeq
where the symbols have the usual meaning. 

By definition, the dual superconformal generators act on the dual coordinates $\{ x_i^{\al\be}, \te_i^\al \}$ in the standard way, 
namely, as the following differential operators 
\beq
\begin{split}\label{dscgen}
P_{\al\be} &= \sum_{i=1}^n \frac{\pr}{\pr x_i^{\al\be}}, \quad D = - \sum_{i=1}^n \left( x_i^{\al\be} \frac{\pr}{\pr x_i^{\al\be}} + \half \te_i^\al \frac{\pr}{\pr \te_i^\al} \right), \\
Q_\al &= \sum_{i=1}^n \frac{\pr}{\pr \te_i^\al} , \quad \bar Q_\al = \sum_{i=1}^n \te_i^\be \frac{\pr}{\pr x_i^{\be\al}} , \\
M_{\al\be} & = \sum_{i=1}^n \left( x_{i\al}^{~~\ga} \frac{\pr}{\pr x_{i}^{\ga\be}} + \half \te_{i\al} \frac{\pr}{\pr \te_{i}^{\be}} \right) , \quad R = \sum_{i=1}^n \te_i^\al \frac{\pr}{\pr \te_i^\al} 
\end{split}
\eeq
To define the action of $\left( K_{\al\be}, S_\al, \bar S_\al \right)$ we make use of the standard technique of {\it Inversion} operation that acts on the dual coordinates as follows 
\beq\label{invDC}
I \left[ x_i^{\al\be} \right] = \f{x_i^{\al\be}}{x_i^2}, \quad  I \left[ \te_i^{\al} \right] = \f{ x_i^{\al\be} \te_{i\be} }{x_i^2}
\eeq
Using the Inversion operation, $\left( K_{\al\be}, S_\al, \bar S_\al \right)$ can be implemented as follows
\beq
\begin{split}\label{KSbS}
K_{\al\be} = I P_{\al\be} I , \quad S_\al = I Q_\alpha I , \quad \bar S_\al = I \bar Q_\al I. 
\end{split}
\eeq

In the next subsection we show that the amplitude \eqref{eq:super4pt} is invariant under the action of dual superconformal symmetry generated by the operators defined in this section.

%%%%%%%%%%%%%%%%%%%%%%%%%%%%%%%%%%%%%%%%%%
\subsection{Dual superconformal symmetry of four point amplitude }\label{DSC4ptamp}
The super-amplitude \eqref{eq:super4pt}, rewritten in terms of the dual coordinates using \eqref{Ceqns}, takes the following form
\beq
\begin{split}\label{ampDC}
{\cal A}_4 & = \sqrt{\f{x_{1,3}^2}{x_{2,4}^2}}\delta^{(3)}(x_1 - x_5)\delta^{(2)}(\theta_1 - \theta_5) \\
\end{split}
\eeq
where, in defining ${\cal A}_4$, we have stripped off the coupling constant dependent factor that appears in \eqref{eq:super4pt} and \eqref{loopans1}.
 
In this form, the invariance of ${\cal A}_4$ under super-translations and Lorentz rotations generated by $\{ P_{\al\be}, Q_\al, M_{\al\be} \}$ is obvious, since ${\cal A}_4$ is only a function of squared differences of the dual coordinates\footnote{The delta functions are also clearly translationally and rotationally invariant.}. Further, ${\cal A}_4$ transforms as an eigenfunction of weight $+4$ and $+2$ under $D$ and $R$ respectively. The invariance under $\bar Q_\al$, though not readily obvious, follows from straightforward algebra that we outline in appendix \eqref{barQinv}. 

In the next two subsections we will show the invariance of ${\cal A}_4$ under $K^{\alpha\beta}$ and $(S_\alpha , \bar{S}_\alpha)$ respectively using the action of Inversion operation on dual coordinates \eqref{KSbS}. 

%%%%%%%%%%%%%%%%%%%%%%%%%%%%%%%%%%%%%%%%
\subsubsection{Invariance under \texorpdfstring{$K^{\alpha\beta}$}{K}}
Since we will be using Inversion on ${\cal A}_4$, we would need the transformation of the $\delta$-functions under Inversions. Using the definitions
\beq 
\int d^3x ~\delta^{(3)}(x) =1 , \quad \delta^{(2)}(\theta) = \te^\al \te_\al 
\eeq
and the transformation \eqref{invDC}, it follows that 
\beq\begin{split}
I \left[ \delta^{(3)}(x) \right] = x^6 \delta^{(3)}(x), ~~
I \left[ \delta^{(2)}(\te) \right] = \frac{1}{x^2} \delta^{(2)}(\te) .
\end{split}\eeq
This implies the following transformation property for the product of these $\delta$-functions 
\beq
\begin{split}\label{scld}
& I \left[ \delta^{(3)}(x_1 - x_5) \delta^{(2)}(\theta_1 - \theta_5) \right] \\
        & \qquad = x_1^4 ~\delta^{(3)}(x_1 - x_5) \delta^{(2)}(\theta_1 - \theta_5).
\end{split}\eeq
This should be contrasted with the case of 4d ${\cal N}=4$ SYM  and ${\cal N}=6$ ABJM, where the similar product of $\delta$-functions turns out to be invariant under dual Inversions. Below we show that in spite of this non-invariance, the action of $K_{\al\be}$ and $(S_\al, \bar S_\al)$ on ${\cal A}_4$ turns out to have a particularly simple form consistent with dual superconformal invariance. 

To show this let us apply the $K_{\al\be}$ in \eqref{KSbS} on our amplitude ${\cal A}_4$ in \eqref{ampDC}, we get
\begin{align}\label{Deltavlu}
K_{\alpha\beta} \left[ {\cal A}_4 \right]
        =  I P_{\alpha\beta} I \left[ {\cal A}_4 \right]= I \sum_{i=1}^4 \pr_{i\alpha\beta} \left[ x_1^4 \sqrt{\frac{x_2^2 x_4^2}{x_1^2 x_3^2}} ~ {\cal A}_4 \right] 
\end{align}
Now, using the invariance of  ${\cal A}_4$ under $P_{\al\be}$ and \eqref{invDC} we have 
\beq
\begin{split}
K^{\alpha\beta} \left[ {\cal A}_4 \right] =&  I \left[ - \frac{1}{2} x_1^4 \sqrt{\frac{x_2^2 x_4^2}{x_1^2 x_3^2}} \left( 3 \frac{x_1^{\alpha\beta}}{x_1^2} 
          + \frac{x_2^{\alpha\beta}}{x_2^2}+ \frac{x_4^{\alpha\beta}}{x_4^2} 
          - \frac{x_3^{\alpha\beta}}{x_3^2} \right) {\cal A}_4 \right]  \\
& =  - \frac{1}{2} \left( 3 x_1^{\alpha\beta} + x_2^{\alpha\beta} + x_4^{\alpha\beta} - x_3^{\alpha\beta}  \right){\cal A}_4 \\
& =  - \frac{1}{2} \left( \sum_{j=1}^{4} \Delta_j x_j^{\alpha\beta} \right) {\cal A}_4
            \qquad \textrm{w}/  \quad \{ \Delta_j \} = \{ 3,1,-1,1 \} \\ 
\end{split}
\eeq
which can be rewritten as 
\beq\label{dscKinv}
 \tilde{K}^{\alpha\beta} {\cal A}^{(4)}_{CS}  = 
        \left( K^{\alpha\beta} + \frac{1}{2} \sum_{j=1}^{4} \Delta_j x_j^{\alpha\beta} \right) {\cal A}^{(4)}_{CS} = 0 \
\eeq

This establishes the invariance of ${\cal A}_4$ under the shifted dual special conformal generator $\tilde{K}_{\al\be}$ as in the case of ABJM and ${\cal N}=4$ SYM theories. Let us note that as opposed to the case of ${\cal N}=6$ ABJM and 4d ${\mathcal N}=4$ SYM) where one gets $\{ \Delta_j \} = \{1,1,1,1\}$, we get different values, namely $\{ \Delta_j \} = \{ 3,1,-1,1 \}$. The differences in the weights is associated with the reduction of supersymmetry.

%%%%%%%%%%%%%%%%%%%%%%%%%%%%%%%%%%%%%%%%%%
\subsubsection{Invariance under \texorpdfstring{$S_{\alpha}$ and $\bar{S}_\alpha$}{S and S-bar}}
To show the invariance under dual conformal supercharges $S_\al$ and $\bar S_\al$, we follow the same procedure as in the previous subsection. Let us first look at the action of $S_\al$. The invariance under $S_\alpha$ works in a rather trivial fashion
\beq
\begin{split}\label{dscSinv}
S_\alpha \left[ {\cal A}_4 \right] &= I Q_\alpha I \left[ {\cal A}_4 \right] = I Q_\alpha \left[ x_1^4 \sqrt{\frac{x_2^2 x_4^2}{x_1^2 x_3^2}} ~ {\cal A}_4 \right]= 0 .
\end{split}
\eeq
since $Q_\al$ annihilates all the terms inside the bracket. 

The invariance under $\bar{S}_\alpha$, on the other hand, works non trivially and in a manner very similar 
to that under $K_{\alpha\beta}$
\beq
\begin{split}
\bar{S}_\alpha \left[ {\cal A}_4 \right] &=  I \bar{Q}_\alpha I \left[ {\cal A}_4 \right] = I  \left[ \left( \sum_{i=1}^4 \te_i^\be \pr_{i\be\al} \right) x_1^4 \sqrt{\frac{x_2^2 x_4^2}{x_1^2 x_3^2}} ~ {\cal A}_4 \right] \\
  & =- \frac{1}{2}\left[ I \left( 3 \frac{\theta_1^\beta x_{1\beta\alpha}}{x_1^2} 
             + \frac{\theta_2^\beta x_{2\beta\alpha}}{x_2^2} - \frac{\theta_3^\beta x_{3\beta\alpha}}{x_3^2} 
             + \frac{\theta_4^\beta x_{4\beta\alpha}}{x_4^2}\right)  \right]{\cal A}_4\\
 & = - \frac{1}{2} \left( \sum_{j=1}^{4} \Delta_j \theta_{j\alpha} \right)  {\cal A}_4 
          \qquad  \textrm{w}/ \quad  \{ \Delta_j \} = \{ 3,1,-1,1 \} \\
\end{split}
\eeq
which can be rewritten as
\beq\label{dscBSinv}
 \tilde{\bar{S}}_\alpha  \left[ {\cal A}_4 \right] 
       = \left(  \bar{S}_\alpha + \frac{1}{2} \left( \sum_{j=1}^{4} \Delta_j \theta_{j\alpha} \right) \right) 
         \left[ {\cal A}_4 \right] = 0 
\eeq
This establishes the invariance of the amplitude ${\cal A}_4$ under the shifted generators $\tilde{\bar{S_\al}}$. Notice that in this case as well the required shift in $\bar S_\al$ has exactly the same form as $\{ \Delta_j \}$ for $K_{\al\be}$. Together \eqref{dscKinv}, \eqref{dscSinv} and \eqref{dscBSinv} prove dual $osp(2|4)$ superconformal invariance of our scattering amplitude \eqref{eq:super4pt}.

%%%%%%%%%%%%%%%%%%%%%%%%%%%%%%%%%%%%%%%%%%%%%%%%%%%%%%%%%%
\subsection{Dual-Superconformal invariance at tree and loop level  at leading order in the $\frac{1}{N}$ expansion}
The computations in the previous subsection show that the tree level scattering amplitude, related to the ${\cal A}_4$ via 
\beq\label{treea}
T_{\rm{tree}} = \frac{4\pi}{\kappa} \mathcal A_4.
\eeq
is dual superconformal invariant. Further, as discussed in section \eqref{scat}, in the 't Hooft large $N$ limit the all loop $2\rightarrow2$ scattering amplitude in this theory, \eqref{loopans1}, is tree level exact except in the anyonic channel where it gets renormalized by simple function of 't Hooft coupling ($\lambda$). Since the momentum dependence of the all loop amplitude and tree level amplitude are identical to that of $\mathcal A_4,$ it follows that the dual superconformal invariance is exact symmetry to all loops in our theory.  

The dual superconformal invariance shown above provides a possible explanation for the simplicity  {\color{blue} of the large $N$} exact $2\to 2$ S-matrix in this theory. In the next section, we explore this question in the reverse direction i.e. how much does exact loop level dual superconformal invariance constrains the S-matrix.

%%%%%%%%%%%%%%%%%%%%%%%%%%%%%%%%%%%%%%%%%%%%%
\section{Constraining  \texorpdfstring{$2\rightarrow2$}{2-to-2} scattering amplitude in CS matter theory based on symmetry}\label{sec:ulta}
So far, we have seen that the all loop exact result for the scattering amplitude exhibits dual superconformal invariance. In this section, we reverse the argument and ask, if it is possible to fix the amplitude based on general symmetry principles. We show that dual superconformal invariance, along with on-shell condition, completely fixes the momentum dependence of the amplitude. Afterwards we further give constraints on the coupling constant dependent piece using a combination of parity invariance, unitarity and self duality of our theory. To start with, we demand that for 2-to-2 scattering amplitude, which is same as a 4-point function in the dual coordinates, the dual super conformal invariance is an exact symmetry to all loops in the planar limit.
 
To implement this, let us compute four point function of operators ${\mathcal O_i}(x_i)$ of scaling dimension $\Delta_i$ in dual space. These operators are positioned at  $x_i$ such that $\left( x_{i+1}-x_{i }\right)^2=0$. The dual conformal invariance implies 
\begin{equation}\label{4pt}
\begin{split}
&\langle {\mathcal O_1}(x_1){\mathcal O_2}(x_2){\mathcal O_3}(x_3){\mathcal O_4}(x_4) \rangle\\
&=\frac{1}{x_{12}^{\Delta_1+\Delta_2} x_{34}^{\Delta_3+\Delta_4}}\left(\frac{x_{24}}{x_{14}} \right)^{\Delta_1-\Delta_2}\left(\frac{x_{14}}{x_{13}} \right)^{\Delta_3-\Delta_4}f(u,v,\kappa,\lambda)
\end{split}
\end{equation}
where $u,v$ are conformal cross ratios defined by
\begin{equation}
u=\frac{x_{12}^2 x_{34}^2}{x_{13}^2 x_{24}^2},~~v=\frac{x_{14}^2 x_{23}^2}{x_{13}^2 x_{24}^2}.
\end{equation}
The function $f(u,v,\kappa,\lambda)$ is a function of cross ratios as well as coupling constants.
Let us notice that on-shell both the cross ratios $u,v$ vanish as
\begin{equation}\label{uv0}
\begin{split}
&~~~~~~~~x_{12}^2=p_1^2=0,~~x_{14}^2=p_4^2=0,\\
&~~~~~~~~x_{34}^2=p_3^2=0,~~x_{23}^2=p_2^2=0\\
&\implies u=0,~~~v=0.
\end{split}
\end{equation}
This immediately implies 
\begin{equation}\label{onshelldual}
f(u,v,\kappa,\lambda)= g(\kappa,\lambda)
\end{equation}
where the function $g$ is independent of $x_{ij}$ and hence independent of momentum.\footnote{In momentum space the on-shell S-matrix is a function of two of the $s,t,u$ variables. In general at loop level, one gets complicated functions (having branch cuts, poles) of Mandelstam variables. However, dual conformal invariance implies momentum  dependence is rather trivial.} Hence, the most general expression for the 4-point amplitude of massless particles that is invariant under dual conformal invariance is given by,
\begin{equation}
	\langle {\mathcal O_1}(x_1){\mathcal O_2}(x_2){\mathcal O_3}(x_3){\mathcal O_4}(x_4) \rangle
=\frac{1}{x_{24}^{\Delta} ~x_{13}^{\Delta'}} \, g(\ka,\la). ~\footnotemark
\end{equation}%
\footnotetext{The scattering matrix is related to this correlation function by, $T_4 = \frac{1}{x_{24}^{\Delta} ~x_{13}^{\Delta'}} \, g(\ka,\la) ~ \de(x_1-x_5) \, \de(\theta_1-\theta_5) = g(\ka,\la) \mathcal A_4$.}%
Additionally, using dimensional analysis in real space (as opposed to the dual space)\footnote{The mass-dimension of the dual coordinates $x_i$ is same as the mass-dimension of real momentum, $\left[ x_i \right] = \left[ p_i \right] = 1$} one can show that the \emph{mass-dimension} of $\left[ \mathcal A_4 \right] = -2$, which implies that $\Delta'=-\Delta$. This follows from computing the scattering cross-section using $\mathcal A_4$. Since the mass-dimension of scattering cross-section is always $(-2)$, same as that of area, one can use that to fix the dimensions of $\mathcal A_4$. In general, one would expect that $\Delta$ can get corrected by loop-diagrams and need not be same as the tree-level answer. However, in \cite{dscinv}, we show using Yangian invariance that $\De=1$, which is the same as the tree-level answer.\footnote {One possible way $\Delta$ can get corrected is due to IR divergences. In all the computations of scattering amplitudes in Chern-Simons theories with fundamental matter done so far \cite{Inbasekar:2015tsa, Jain:2014nza, Yokoyama:2016sbx}, no IR divergences were observed. It is not clear why this is so, since it is known that IR divergences do appear in ABJM theories. For the purposes of this paper, we proceed with the assumption that there are no IR divergences in the theories in consideration. }
With these assumptions, we see that the momentum dependence of the scattering amplitude is completely fixed. In \autoref{subsec:parity-unitarity-duality}, we show that constraints imposed by duality, parity and unitarity are not sufficient to fix the coupling constant either.

\subsection{Constraining coupling constant (\texorpdfstring{$\lambda$}{lambda}) dependence of scattering amplitude}\label{subsec:parity-unitarity-duality}
The only other way the S-matrix can get corrected is by $g$ becoming a function of the coupling constant. We ask, is it possible to determine $g(\kappa,\lambda)$ in \eqref{onshelldual} based on some general symmetry principles? In this section, we will study the constraints imposed by following symmetries on the scattering amplitude: 
\begin{itemize}
\item Parity invariance
\item Unitarity
\item Duality.
\end{itemize}

\paragraph{Parity constraints on the function $g(\ka,\la)$:} Let us start by looking at the transformation properties of the S-matrix under parity transformation. From the definition 
\beq
\begin{split}
S(\{p_i\}) &= {\cal I}(\{p_i\}) + i T(\{p_i\}) \\ 
&=  \langle \Phi(p_4) \Phi(p_3) \Phi^\dagger(p_2) \Phi^\dagger(p_1) \rangle , \\
\textrm{where} & \quad \Phi(p)  = a(p) + \eta \al(p) , \\ 
& \quad  \Phi^\dagger(p) = \eta^\dagger a^\dagger(p) + \al^\dagger(p) \\
\end{split}
\eeq
and the action of parity on the on-shell superfield $\Phi(p)$
\beq 
{\cal P} [\Phi(p)] = \Phi(-p)
\eeq
it follows that 
\beq
\begin{split}\label{parityST}
& {\cal P} \left[ S(\{p_i\}) \right] = S(\{-p_i\}) \\
 \Rightarrow \quad & {\cal P} \left[ T(\{p_i\}) \right] = T(\{-p_i\}) .
\end{split}
\eeq

Now, as we have seen earlier, exact dual conformal invariance fixes the all loop amplitude to be proportional to tree level result \eqref{loopans1}, \eqref{loopans2} with only an overall coupling dependent factor, $g(\kappa, \la)$\footnote{$\la$ is the 't Hooft coupling defined as $\lambda=N/\kappa$, not to be confused with similar notation used for the on-shell spinor variables $\{\lambda_{i\al}\}$.}, unfixed. To determine the constraint of parity on the function $g(\kappa, \la)$, let us first look at the action of parity on the coupling independent part of tree amplitude, namely \eqref{ampDC}. The parity transformation of the constituent elements of ${\cal A}_4$ are as follows 
\begin{equation} 
\begin{split}
{\cal P} [ ( p_{i}^{\al\be}, \lambda_{i\al}, \eta_i ) ] &= ( - p_{i}^{\al\be}, i\lambda_{i\al},  \eta_i ) \\
  \Rightarrow \quad {\cal P} [Q_\al] &= i Q_\al.
\end{split}\end{equation}
This give us that ${\cal A}_4$ is parity odd \footnote{Recall, $\mathcal A_4 = \frac{1}{x_{24}^{\Delta} ~x_{13}^{\Delta'}} ~ \de(x_1-x_5) \, \de(\theta_1-\theta_5)$.}
\beq
\begin{split}\label{A4parityodd}
{\cal P} \left[ {\cal A}_4 \right] & = - {\cal A}_4. 
\end{split}
\eeq
Since the momentum dependence is now already completely fixed, \eqref{parityST} implies that the exact S-matrix, and hence $T$, is parity invariant as it is an even function of the external momenta. 
\beq\label{Tparityinv}
{\cal P} [T] = T .
\eeq
Further, using \eqref{Tparityinv} , \eqref{A4parityodd} together with 
\beq
T = g(\kappa,\lambda) {\cal A}_4 \quad , \quad {\cal P}[(\kappa, \la)] = ( - \kappa, - \la) 
\eeq
implies that the function $g(\kappa, \la)$ should be parity odd 
\beq\label{partyco}
{\cal P} [g(\kappa, \la)] = g( - \kappa, - \la) = - g(\kappa, \la).
\eeq

To proceed further, let us distinguish between anyonic and non-anyonic channels. As explained in Eq.(2.41) of \cite{Jain:2014nza}, the amplitude in the non-anyonic channel is of order 
$\frac{1}{\kappa}$ and in the anyonic channel, it is of order $\lambda$.
In the non-anyonic channel, we expect
\beq\label{eq:pert-const-nonan}
\begin{split}
& T_{tree}\sim {\mathcal O} \left( \frac{1}{\kappa} \right), \quad T_{n-loop}\sim {\mathcal O} \left( \frac{\lambda^n}{\kappa} \right). 
\end{split}
\eeq
For anyonic channel we expect to get
\begin{equation}
T_{tree}\sim {\mathcal O}(\lambda), \quad T_{n-loop} \sim {\mathcal O}(\lambda^{n+1}).
\end{equation}
We observe that, at odd number of loops, \eqref{partyco} is violated and hence we conclude that one loop as well as any odd loop answer has to vanish. This is consistent with Eq.\eqref{loopans1} and \eqref{loopans2}.

\paragraph{Duality constraints on $g(\ka,\la)$:} We have seen that dual conformal invariance fixes the momentum dependence completely for four point function but leaves the function $\Delta(\lambda),~g(\kappa,\lambda)$ unfixed.  Let us further assume that there is no IR divergence and $\Delta(\lambda)=\Delta(\lambda=0)$. Furthermore, self duality of ${\mathcal N}=2$ under \eqref{dualityst} implies
\begin{equation}\label{duality2}
\begin{split}
g(\kappa,\lambda)&=\pm g(-\kappa,{\la-\text{sgn}(\lambda)}).
\end{split}\end{equation}
where the last equality is valid up to a sign. To fix the sign ambiguity, we take a special point $\lambda=\frac{1}{2}$ which implies
\begin{equation}\label{midpnt}
g(\kappa,\frac{1}{2})=\pm g(-\kappa,-\frac{1}{2}).
\end{equation}
Now comparing \eqref{midpnt} with  \eqref{partyco}, we immediately get 
\begin{equation}\label{duality1}
\begin{split}
g(\kappa,\lambda)&=- g(-\kappa,{\la-\text{sgn}(\lambda)}).
\end{split}\end{equation}
{\bf{Duality constraint in non-anyonic channel:}}
In the non-anyonic channel, following \eqref{eq:pert-const-nonan}, we have
\begin{equation}
g(\kappa,\lambda)=\frac{4\pi}{\kappa}f(\lambda)
\end{equation}
at the tree level, $f(\lambda=0)=1$ which matches with the tree level answer in \eqref{loopans1}.
The function $f(\lambda)$ satisfies
\begin{equation}\label{fn1g}
\begin{split}
f(\lambda)&=f(-\lambda)\\
f(\lambda)&=f({\la-\text{sgn}(\lambda)})\\
f(0)&=1
\end{split}\end{equation}
Equation \eqref{fn1g} implies
\begin{equation}\label{period}
f(1)=f(-1)=f(0)=1.
\end{equation}
It is clear that fixing the value of a function at three-points is not sufficient to constraint it completely. Hence, we deduce that duality constraint doesn't fix the scattering amplitude in the non-anyonic channel completely.
%We Fourier transform the function $f(\lambda)$ and using \eqref{fn1g} we get
%\begin{equation}
%f(\lambda)=\sum_{0}^{\infty} f_{m}\cos\left(\pi \lambda m\right).
%\end{equation}
%Using equation \eqref{period} we get
%\begin{equation}\label{prd1}
%\begin{split}
%f(0)&=\sum_{-\infty}^{\infty} f_{m}=1\\
%f(1)=f(-1)&=\sum_{-\infty}^{\infty} f_{m}(-1)^{m}=1\\
%\end{split}\end{equation}

\paragraph{Unitarity constraint on function $f(\lambda)$ in non-anyonic channel:}
The unitarity equation for the S-matrix is given by
\begin{equation}\label{untnan}
SS^\dagger =1\implies T-T^{\dagger}=i ~T ~T^{\dagger}. 
\end{equation}
In the case of the non-anyonic channel, the scattering matrix $T\sim {\mathcal O}( \frac{1}{\kappa})$ (see \eqref{loopans1}) and hence $T T^{\dagger}\sim {\mathcal O}(\frac{1}{\kappa^2})$.
Therefore, we get no more further constraint on the coupling constant dependence piece. So we conclude that, general symmetry principles can't fix the coupling constant dependence fully.
 
\paragraph{Duality and unitarity constraint in anyonic channel:} A similar analysis can be carried out in the anyonic channel and in this case as well, these general principles are not sufficient enough to fix the coupling constant dependence. However, in the case of anyonic channel unitarity equation \eqref{untnan} gives a nontrivial constraint on the amplitude. This is easy to understand. In the anyonic channel, $T\sim {\mathcal O}(\lambda)$ and hence order one, as compared to non-anyonic channel where amplitude is order $\frac{1}{N}$ suppressed.
So $T T^{\dagger}\sim {\mathcal O}(1)$. Although unitarity gives non-trivial constant on the amplitude, one can check that these constraints are  not sufficient to fix the amplitude.

In \autoref{app:explicit}, we show that for ${\mathcal N}=2$ CS matter theory, at large $N$, $\Delta$ is coupling independent and $g$ is a simple function of coupling constant, thus  the $2\to 2$ scattering amplitude of the ${\mathcal N}=2$ CS matter theory exhibits dual super conformal invariance at all loops.

%%%%%%%%%%%%%%%%%%%%%%%%%%%%%%%%%%%%%%%%%%%%%%%%
\section{Discussion}\label{sec:diss}
In this paper we have established an all loop exact dual super conformal invariance of $2\to 2$ scattering amplitude in ${\mathcal N}=2$ Chern-Simons theory with fundamental chiral multiplet, which provides an explanation for the remarkable simplicity of these amplitudes. The analysis in this paper opens up a variety of exciting directions for future research, which we briefly discuss below.

In previous known examples, the existence of dual conformal invariance turns out to be related to the Wilson loop - Scattering amplitude duality, see for e.g. in 4d ${\mathcal N}=4$ SYM \cite{Drummond:2008vq, Alday:2010zy}. It would be very interesting to see if this also the case for our theory. A cursory look at the relevant set of diagrams shows that, at large $N$, they are exactly the same as in pure Chern-Simons theory. This indicates that an all loop exact computation is not out of reach. We plan to report on this in near future. 

Another natural question to ask is, if the scattering amplitude in our theory also enjoys Yangian symmetry \cite{Drummond:2009fd, Bargheer:2010hn, Bargheer:2011mm} observed in other known examples of theories with dual superconformal invariant amplitudes. In \cite{dscinv}, we show that this is indeed the case. Since Yangian is an infinite dimensional symmetry, its presence indicates elements of integrability in the theory and has lead to remarkable developments like the Orthogonal Grassmanian \cite{ArkaniHamed:2012nw, Huang:2013owa, Huang:2014xza} representation for amplitudes. 

It would also be very interesting to see if dual conformal invariance, or perhaps some anomalous version, exist for non-supersymmetric counterparts of the theory studied in this paper. The scattering amplitudes in the non-supersymmetric case were computed in \cite{Jain:2014nza} and, though the amplitudes do get non-trivial renormalisation at loop levels,\footnote{tree-level amplitudes in non-supersymmetric theories were also computed in \cite{Inbasekar:2017ieo} using BCFW recursion relations.} the results still take reasonably simple form and hence one might expect existence of dual conformal invariance. In the non-relativistic limit, the scattering amplitudes in these Chern-Simons matter theories reproduce the Aharonov-Bohm scattering, as was shown in \cite{Jain:2014nza,Inbasekar:2015tsa}. It would be quite remarkable if some non-trivial component of dual (super) conformal symmetry of the relativistic amplitudes persist to the non-relativistic limit i.e. Aharonov-Bohm scattering. This would require us to go beyond massless limit studied in this paper. 

One of the puzzling features of scattering amplitudes in ABJM theory is the two loop unitarity puzzle \cite{Agarwal:2008pu}, namely that the one-loop amplitude vanishes where as two loop amplitude is non vanishing. Dual super-conformal invariance was used extensively to make this puzzle sharp \cite{Bianchi:2014iia}. A similar problem with unitarity arises in Chern-Simons theory coupled to fundamental matter. However, it was remarkably resolved by giving up the standard analyticity properties of the scattering amplitudes \cite{Jain:2014nza, Inbasekar:2015tsa, Yokoyama:2016sbx}, resulting in modification of crossing symmetry rules. We suspect that the resolution of unitarity puzzle for ABJM scattering might also work in a similar fashion. Study of scattering in Chern-Simons matter theories with higher supersymmetry ${\cal N}=4,6$, in the vector model limit, could shed light on these issues. Quite possibly, dual superconformal invariance may exist for these theories as well, leading to similar puzzle with unitarity. Taking a lesson from the resolution of the puzzle in simpler theories with fundamental matter might give us useful insights into the unitarity puzzle in ABJM theory. 

For the case of ${\cal N}=4$ SYM, dual superconformal symmetry was explained in \cite{Berkovits:2008ic} as a certain T-duality involving bosonic and fermionic isometries in the dual $AdS_5 \times S^5$ type IIB string theory. Attempts at finding a similar explanation for the dual superconformal symmetry of ABJM amplitudes on the other hand have encountered difficulties \cite{Adam:2009kt, Adam:2010hh, Bakhmatov:2010fp, Dekel:2011qw}. Given that the holographic dual of theory considered in this paper are Vasiliev higher spin theories, it would be interesting to see as to what does the dual superconformal symmetry, assuming its a exact symmetry for all higher point amplitudes, imply for the holographic Vasiliev duals? We leave these exciting possibilities for future research.

\begin{figure}[h]
\begin{center}
\includegraphics[width=6cm,height=2.5cm]{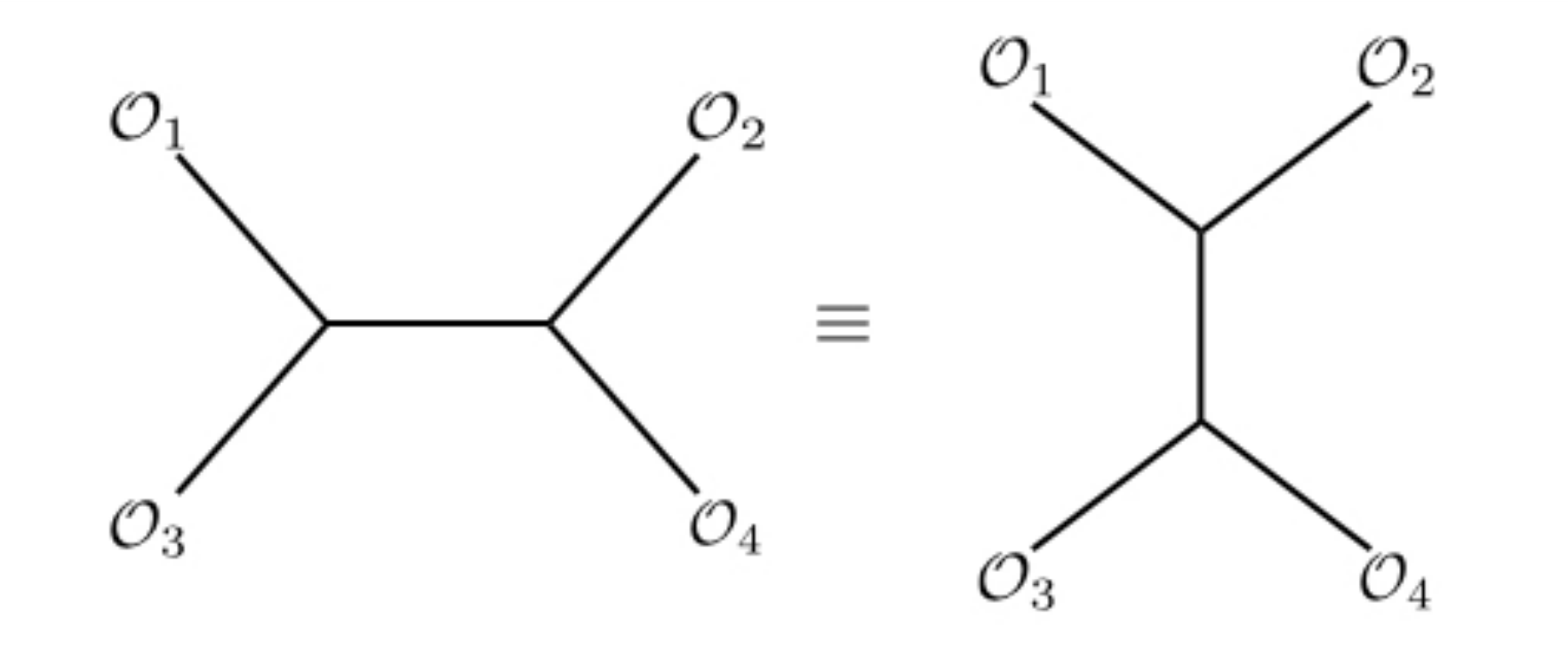}
  \caption{Crossing relations for the scattering amplitude in dual space.}\label{crossing}
 \end{center}
\end{figure}

Lastly, following the discussion of \autoref{sec:cft}, it would also be very interesting to understand the scattering amplitude as a 4-point function in a free field theory from the  cross channel OPE expansion. In the cross channel, \autoref{crossing},
double trace operators are exchanged. To reproduce the identity contribution from the other channel, one would generally require all double trace operators with scaling dimensions
\begin{equation}\begin{split}
\Delta_{Double~trace}&=\Delta_1+\Delta_2+l ~ =-\frac{1}{2}+\frac{1}{2}+l ~ =l
\end{split}\end{equation} 
where $l$ is the spin of  double trace operator.
It would be interesting to explicitly show that the same result is reproduced in the other channel. Understanding OPE structure properly might help us compute higher point functions using the conformal symmetry.

Moreover, for a general $n-point$ function the number of independent cross ratios are given by $\frac{n(n-3)}{2}$. For a six point function this gives 9 cross ratios, out of which 6 of them vanish on-shell and the remaining three can be written as \cite{Henn}
\begin{equation}
u=\frac{x_{13}^2 x_{46}^2}{x_{14}^2 x_{36}^2},~~v=\frac{x_{24}^2 x_{15}^2}{x_{25}^2 x_{14}^2},~~,w=\frac{x_{35}^2 x_{26}^2}{x_{36}^2 x_{25}^2}.
\end{equation}

So we see that, if the six point amplitude is invariant under dual conformal symmetry, it can be determined up to a function of three variables $f_6(u,v,w,\kappa,\lambda)$. So we see, unlike the four point function, the six point amplitude can get renormalized by a non-trivial function of momentum.  We leave the determination of $f(u,v,w,\kappa,\lambda)$ for future work. It would be interesting to see, if the full six point or higher point\footnote{For an eight point amplitude, one obvious way to proceed is to 
wick contract pairwise all the operators with same scaling dimension. It would be interesting to see if this way of thinking reproduces the correct loop level result consistent with unitarity.} amplitude can be obtained from the OPE expansion discussed above.

\vspace{5mm}

%%%%%%%%%%%%%%%%%%%%%%%%%%%%%%%%%%%%%%%%%%%%%%%%%%%%%%
\textbf{Acknowledgements:} We thank R.Gopakumar and S. Minwalla for very useful and critical discussions which helped us clarify many aspect of the paper. SJ would like to thank A. Thalapillil for enjoyable discussions. The work of KI was supported in part by a center of excellence supported by the Israel Science Foundation (grant number 1989/14), the US-Israel bi-national fund (BSF) grant number 2012383 and the Germany Israel bi-national fund GIF grant number I-244-303.7-2013. S.J. would like to thank TIFR for hospitality at various stages of the work. The work of P.N. and R.S. was supported partly by Infosys Endowment for the study of the Quantum Structure of Space Time. P.N. also acknowledges support from the College of Arts and Sciences of the University of Kentucky and Indo-Israel grant of S. Minwalla. T.S. would like to thank TIFR for hospitality while this work was in progress. K.I, P.N, R.S and V.U would like to thank IISER Pune for hospitality during the course of this work. We would like to thank Ofer Aharony, Tom Hartman,
Yu-tin Huang and Shimon Yankielowicz for helpful comments on the draft.  SM acknowledges support from a CSIR NET fellowship. TN acknowledges support from a UGC NET fellowship. Finally, we would also like to thank people of India for their steady support in basic research.

%%%%%%%%%%%%%%%%%%%%%%%%%%%%%%%%%%%%%%%%%%%%%
\appendix

%%%%%%%%%%%%%%%%%%%%%%%%%%%%%%%%%%%%%%%%%%%%%
\section{Notations and Conventions}\label{notconv}
In this appendix we list the conventions and notation used in the main text. 
\begin{itemize}
\item We work in the Minkowski metric with signature $\{-,+,+\}$. 
\item Spinor indices are raised and lowered using anti-symmetric $\epsilon$ symbol as follow
\beq
\begin{split}\label{}
& v^\al = \epsilon^{\al\be} v_\be, \quad v_\al = \epsilon_{\al\be} v^\be , 
           \quad \textrm{w/} \quad \epsilon^{12} = - \epsilon^{21} = 1 = - \epsilon_{12} = \epsilon_{21}. \\
& v.u = v^\al u_\al = - v_\al u^\al. 
\end{split}
\eeq
\item Our gamma matrices are
\beq
\ga^o = i\sigma^2, \quad \ga^1 = \sigma^1, \quad \ga^2 = \sigma^3 
\quad \textrm{which satisfy } \quad [\ga^\mu, \ga^\nu ] = \epsilon^{\mu\nu\rho} \ga_\la. 
\eeq
\item Vector and spinor indices can be converted into each other using $\gamma$ matrices as follow 
\beq
\begin{split}
& x_\al^{~\be} = x^\mu \left( \gamma_\mu \right)_\al^{~\be} \quad \Leftrightarrow \quad x^\mu = \half x_\al^{~\be} \left( \ga^\mu \right)_\be^{~\al}. \\ 
& \textrm{Further :}  \quad x^2 \equiv x^\mu x_\mu = - \half x^{\al\be} x_{\al\be}. 
\end{split}
\eeq 
\item We interchangeably use the following notation to represent the spinor helicity variables $\lambda_{i\al}$ 
\beq
| i \rangle = \lambda_{i\al}, \quad \langle ij \rangle = \lambda_i^\al \lambda_{j\al} = - \la_j^\al \la_{i\al} = - \langle ji \rangle
\eeq
\end{itemize}

%%%%%%%%%%%%%%%%%%%%%%%%%%%%%%%%%%%%%%%%%%%%%
\section{Useful Identities}\label{uiden}
In this appendix we list some useful formulae and identities used in the main text of this paper. 
\begin{itemize}
\item The derivative operator with spinor indices acts as follow
\beq\begin{split}
& \pr_{\al\be} x^{\sigma\rho} \equiv \frac{\pr}{\pr x^{\al\be}} x^{\sigma\rho} 
        = \half \left( \delta^\sigma_\al \delta^\rho_\be + \delta^\rho_\al \delta^\sigma_\al \right) , \\
\Rightarrow & \quad \pr_{\al\be} \left( x^2 \right) = \pr_{\al\be} \left( -\frac{1}{2} x^{\ga\delta}x_{\delta \ga} \right)
          = -\delta^{\ga}_{\al}\delta^{\delta}_{\be}x_{\delta \ga} = -x_{\al\be} . \\ 
& 
\end{split}
\eeq
\item The two component spinor helicity variables satisfy the following linear algebra identity :
\beq\label{schID}
\textrm{\underline{Schouten Identity} :} \quad | i \rangle \langle jk \rangle + | j \rangle \langle ki \rangle +| k \rangle \langle ij \rangle = 0 \quad \textrm{for} \quad i\neq j \neq k
\eeq
\item Momentum and super-momentum conservation of four point scattering in  equations \eqref{pqcons} can be used to prove the following simple identities
\beq\begin{split}
\textrm{for} \quad i \neq j \neq k \neq l \quad : \quad 
& \langle ij \rangle^2 = \langle kl \rangle^2, \quad  \langle ij \rangle^2 + \langle jk \rangle^2 + \langle ki \rangle^2 = 0,  \\
& \eta_i \eta_j \langle ij \rangle = \eta_k \eta_l \langle kl \rangle, 
         \quad \eta_i \eta_j \langle ij \rangle + \eta_j \eta_k \langle jk \rangle + \eta_k \eta_i \langle ki \rangle = 0 
\end{split}\eeq
\end{itemize}

%%%%%%%%%%%%%%%%%%%%%%%%%%%%%%%%%%%%%%%%%%%%%%%%%%
\section{\texorpdfstring{$\bar Q_\al$}{Q-bar} invariance of \texorpdfstring{${\cal A}_4$}{A4}}\label{barQinv}
\beq
\begin{split}
\bar Q_\al {\cal A}_4 &= \sum_{i=1}^5 \te_i^\be \pr_{i\al\be} 
                \left[ \sqrt{\frac{x_{13}^2}{x_{24}^2}} \delta^{(3)}(x_{15}) \delta^{(2)}({\te_{15}}) \right] \\
       & =   \sqrt{\frac{x_{13}^2}{x_{24}^2}} ~\delta^{(2)}({\te_{15}}) \te^\be_{15} ~\pr_{1\be\al} \left[ \delta^{(3)}(x_{15}) \right] 
                 + \frac{\delta^{(3)}(x_{15}) \delta^{(2)}({\te_{15}})}{2} \sqrt{ \frac{x^2_{24}}{x^2_{13}} } 
                 \left[ \frac{-\te^\be_{13} x_{13\be\al} }{x^2_{24}} + \frac{ x^2_{13} \te^\be_{24} x_{24\be\al} }{x^4_{24}} \right] \\
\end{split}
\eeq
The first term vanishes as $\te_{15}^\be \delta^{(2)}(\te_{15}) =0$. The second term can be converted to the the on-shell variables and using the Schouten identity \eqref{schID}, takes the form
\beq
\begin{split}
\bar Q_\al {\cal A}_4 & = \frac{ \delta^{(3)}(x_{15}) \delta^{(2)}({\te_{15}}) }{\langle 23 \rangle^2} 
               \bigg[ \eta_1 \langle 23 \rangle + \eta_2 \langle 31 \rangle + \eta_3 \langle 12 \rangle \bigg] | 2 \rangle_\al \\
& = \frac{ \delta^{(3)}(x_{15}) \delta^{(2)}({\te_{15}}) }{\langle 23 \rangle^2} 
               \bigg[ \eta_1 \langle 41 \rangle + \eta_2 \langle 42 \rangle + \eta_3 \langle 43 \rangle \bigg] | 2 \rangle_\al \\
& = \frac{ \delta^{(3)}(x_{15}) \delta^{(2)}({\te_{15}}) }{\langle 23 \rangle^2} 
                \bigg[ \langle 4 | \bigg( \eta_1 | 1 \rangle + \eta_2 | 2 \rangle + \eta_3 | 3 \rangle \bigg) \bigg] | 2 \rangle_\al \\ 
& = \frac{ \delta^{(3)}(x_{15}) \delta^{(2)}({\te_{15}}) }{\langle 23 \rangle^2} 
                \bigg[ - \eta_4 \langle 4 | 4 \rangle \bigg] | 2 \rangle_\al \\
& = 0 
\end{split}
\eeq

%%%%%%%%%%%%%%%%%%%%%%%%%%%%%%%%%%%%%%%%%%%%%%%%%%
\section{Explicit verification}\label{app:explicit}
In \cite{Inbasekar:2015tsa}, the exact $2\rightarrow2$ at all loop was computed and it was noted that the all loop result is the same as tree level except in anyonic channel where it gets renormalized by a simple function of coupling constant. The tree level answer is given by
\begin{equation}
T_{{\rm tree}}=\f{4\pi}{\ka}\sqrt{\f{x_{1,3}^2}{x_{2,4}^2}}\delta^{(3)}(x_1 - x_5)\delta^{(2)}(\theta_1 - \theta_5).
\end{equation}
The all loop answer in the non-anyonic channel is given by
\begin{equation}\label{nonanall}
\begin{split}
 T_{{\rm all-loop}}^{{\rm non-anyonic}} &=T_{{\rm tree}} \\
\implies g(k,\lambda)^{{\rm non-anyonic}}&= \frac{4\pi}{\kappa}.
\end{split}
\end{equation}
In case of anyonic channel we have 
\begin{equation}\label{anall}
\begin{split}
T_{{\rm all-loop}}^{{\rm anyonic}} &=N \frac{\sin(\pi\lambda)}{\pi\lambda}T_{{\rm tree~level}}\\
\implies g(k,\lambda)^{{\rm anyonic}}&= 4\sin(\pi\lambda).
\end{split}
\end{equation}

This immediately implies 
\begin{equation}\label{delg}
\begin{split}
\Delta(\lambda)&=\Delta(\lambda=0)\\
g(\kappa,\lambda)&=-g(-\ka,\la-\text{sgn}(\la))
\end{split}\end{equation}
So we see \eqref{delg} is consistent with duality expectation \eqref{duality1} as well.

Note that, above discussion is in confirmation with the fact that, dual conformal invariance of scattering amplitude is all loop exact.

%%%%%%%%%%%%%%%%%%%%%%%%%%%%%%%%%%%%%%%%%%%%%%%%%%%%%%%%%%%%%%%%%%%%%
%%%%%%%%%%%%%%%%%%%%%%%%%%%%%%%%%%%%%%%%%%%%%%%%%%%%%%%%%%%%%%%%%%%%%
\section{\texorpdfstring{$2\rightarrow2$}{2-to-2} scattering amplitude as free field correlation function  in dual space}\label{sec:cft}
In this subsection we would like to interpret four point function \eqref{ampDC} as a CFT correlator of some operator ${\mathcal O}$. Our motivation is rather observational in nature and we do not claim the existence of a local conformal field theory dual to our original theory.

We begin by noting that, for our case, it is natural to rewrite the scaling dimension of the operators, $\Delta$ obtained in \eqref{Deltavlu}, in the following way \cite{Chicherin:2017cns}
\begin{equation}
\{\Delta_1,\Delta_2,\Delta_3,\Delta_4\}=\frac{1}{2}\{4 - 1,1,-1,1\}
\end{equation}
notice the fact that we write
\begin{equation}
\Delta_1=\f{1}{2}(4-1)=2-{\tilde \Delta}_1,
\end{equation} the factor of $4$ came from momenta and supermomenta conservation in \eqref{scld}.
With this separation we have\footnote{We observe in \eqref{fnlD} that operator $O_1,O_3$ has negative scaling dimension, which in general implies these dual space CFT's are non-unitary.}
\begin{equation}\label{fnlD}
\begin{split}
&{\tilde\Delta}_1=\Delta_3=-\frac{1}{2}\\
&\Delta_2=\Delta_4=\frac{1}{2}
\end{split}\end{equation}
Plugging this in \eqref{4pt} and using \eqref{onshelldual}, we get
\begin{equation}\label{4pta}
\langle {\mathcal O_1}(x_1){\mathcal O_2}(x_2){\mathcal O_3}(x_3){\mathcal O_4}(x_4) \rangle
=g(\kappa,\lambda) \sqrt{\frac{x_{13}^2}{x_{24}^2}}
\end{equation}
which is same as that appeared in \eqref{ampDC} without the delta functions as expected. Note that all the terms $x_{12},x_{14},x_{34}$  whose absolute value vanish, cancel out in \eqref{4pt}. This also serves as a nice consistency check of our computation. 
%So we conclude that, for ${\mathcal N}=2$ theory, at four point function, dual conformal invariance is all loop exact.
\begin{figure}[h]
\begin{center}
\includegraphics[width=4cm,height=2.5cm]{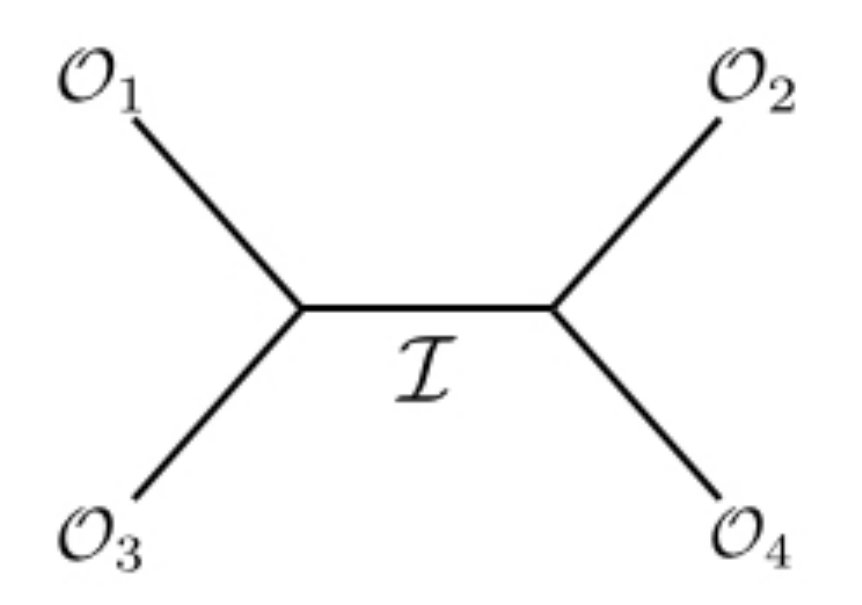}
  \caption{Identity exchange in this channel accounts for full four point function.}\label{identity}
 \end{center}
\end{figure}

We would now like to interpret the function $g(\kappa,\lambda)$ that appears in \eqref{4pta} in terms of an OPE expansion.  We expect that under $u=v=0$, only identity operators will contribute.  Let us start with the channel where operators ${\mathcal O}_2$, ${\mathcal O}_4$ with  $\Delta=\frac{1}{2}$ and operator ${\mathcal O}_1$, ${\mathcal O}_3$ with  $\Delta=-\frac{1}{2}$ are brought together, see Fig.\ref{identity}. Interestingly in this channel we can account for full answer
just by identity exchange which gives
\begin{equation}\label{idntexch}
\begin{split}
&\langle {\mathcal O_1}(x_1){\mathcal O_2}(x_2){\mathcal O_3}(x_3){\mathcal O_4}(x_4) \rangle\\
&=\langle {\mathcal O_1}(x_1){\mathcal O_3}(x_3)\rangle \langle{\mathcal O_2}(x_2){\mathcal O_4}(x_4) \rangle\\
&=c_1 c_2 \sqrt{\frac{x_{13}^2}{x_{24}^2}}
\end{split}
\end{equation}
where 
$c$'s are the coefficient of the two point function of operators $ {\mathcal O}$. This 
Eq.\eqref{idntexch} implies that 
\begin{equation}\label{chk1}
g(\kappa,\lambda)=c_1 c_2.~\footnotemark
\end{equation}
\footnotetext{It would be interesting to understand what these operators mean from CFT point of view and compute their two point functions namely $c_1,c_2$ independently to verify the claim in \eqref{chk1}.}%
So we conclude, as far as $2\rightarrow 2$ amplitude is concerned, results can be interpreted from a free CFT.

\bibliography{dualsc.bib}

\end{document}